\documentclass[twocolumn,epsfig,graphics,mathbbm,nopacs,hyperref]{revtex4-1}
\usepackage{amsmath,amsfonts,amssymb,amsthm,graphics,graphicx,epsfig,times,bbm}
\usepackage{amsmath}
\usepackage{amssymb}
\usepackage{graphics}
\usepackage{graphicx}

\newtheorem{observation}{Observation}

\usepackage{amsfonts}
\usepackage{psfrag}
\usepackage{dsfont}
\usepackage[dvips]{rotating}
\usepackage{epsfig}
\usepackage{pstricks,pst-grad}

\newcommand{\cc}{\mathbbm{C}}

\newcommand{\rr}{\mathbbm{R}}
\newcommand{\id}{\mathbbm{1}}

\begin{document}
\unitlength=1mm

\title{Entangled inputs cannot make imperfect quantum channels perfect}

\date{\today}
\author{
F.\ G.\ S.\ L.\ Brand\~ao$^1$,   J.\ Eisert$^{2,3}$, M.\ Horodecki$^4$, and D.\ Yang$^{5}$}
\affiliation{$^{1}$
Departamento de F\'isica, Universidade Federal de Minas Gerais,
Belo Horizonte 30123-970, Brazil}
\affiliation{$^2$Institute of Physics and Astronomy, University of Potsdam, 14476 Potsdam, Germany}
\affiliation{$^3$Dahlem Center for Complex Quantum Systems, Freie Universit{\"a}t Berlin, 14195 Berlin, Germany}
\affiliation{$^4$Institute of Theoretical Physics and Astrophysics, University of Gda\'nsk, 80-952 Gda\'nsk, Poland}
\affiliation{$^5$Laboratory for Quantum Information, China Jiliang University, Hangzhou, Zhejiang 310018, China}
\begin{abstract}
Entangled inputs can enhance the capacity of quantum channels, this being
one of the consequences of the celebrated result showing the non-additivity of several quantities
relevant for quantum information science. In this work, we answer the
converse question (whether entangled inputs can ever
render noisy quantum channels have maximum capacity) to the
negative: No sophisticated entangled input of any quantum channel
can ever enhance the capacity
to the maximum possible value; a result that holds true for all channels
both for the classical as well as the quantum capacity. This result can hence be seen
as a bound as to how ``non-additive quantum information can be''. As a main result, we
find first
practical and remarkably simple computable single-shot bounds to capacities, related to entanglement
measures. As examples, we discuss the qubit amplitude damping and
identify the first meaningful bound for its classical capacity.
\end{abstract}

\maketitle
How much information can one transmit reliably through a quantum channel such as a
tele-communication fiber? This basic question is,
despite much progress in recent years  \cite{Hastings,Add,Amosov,Shor,RMP},
still
surprisingly wide open. Some suitable encoding and decoding is necessary, needless to say,
but the optimal achievable rates can still not be expressed in a computable closed form.
For classical information, the hope that the single-shot capacity would be sufficient to
arrive at that goal was corroborated by many examples of channels
for which this is in fact true \cite{Add}. Alas, it was finally found to be
unjustified with
the celebrated result \cite{Hastings} on the non-addivity of several quantities that are in the center of
interest in quantum information science \cite{Amosov,Shor,RMP}.
In particular, entangled inputs help and do increase the classical information
capacity. This result showed that the question of finding capacities of quantum channels is more
complicated than what one might have anticipated. In the case of quantum information transmission, a similar situation
has been known to be true already for a long time: in
general one must regularize the single-shot
expression, given by the coherent information, in order to attain the quantum capacity \cite{QNA}.

To contribute to fixing the coordinate system of channel capacities, this insight begs for a
resolution of the
following question: To what extent can entanglement help then? Is the mentioned result
rather an academic observation, manifesting itself in
small violations of additivity in high physical dimensions?
An interesting question in this context is the following: {\it Can suitably
entangled inputs render noisy quantum channels take their maximum
possible capacity or make them even perfect?}
This would be the other extreme, where the non-additivity serves as a resource
to overcome the noisiness of channels.

In this work, we answer this question to the negative: For all quantum channels, no matter how
elaborate the entangled coding over many uses of the channel might be, one can never
achieve the maximum possible capacity if this is not already true on the single-shot level. This
observation holds true both for the classical as well as the quantum capacity.

We show this by introducing new upper bounds to these capacities which can be evaluated
on the single-shot level and are computable, which constitute a main result of this work. We connect questions of capacities to
those of entanglement measures of systems and their environments. These bounds are useful in their own right, which will
be shown by means of an example of amplitude-damping channel.

\begin{figure}
  \begin{center}
 \includegraphics[width=4.7cm]{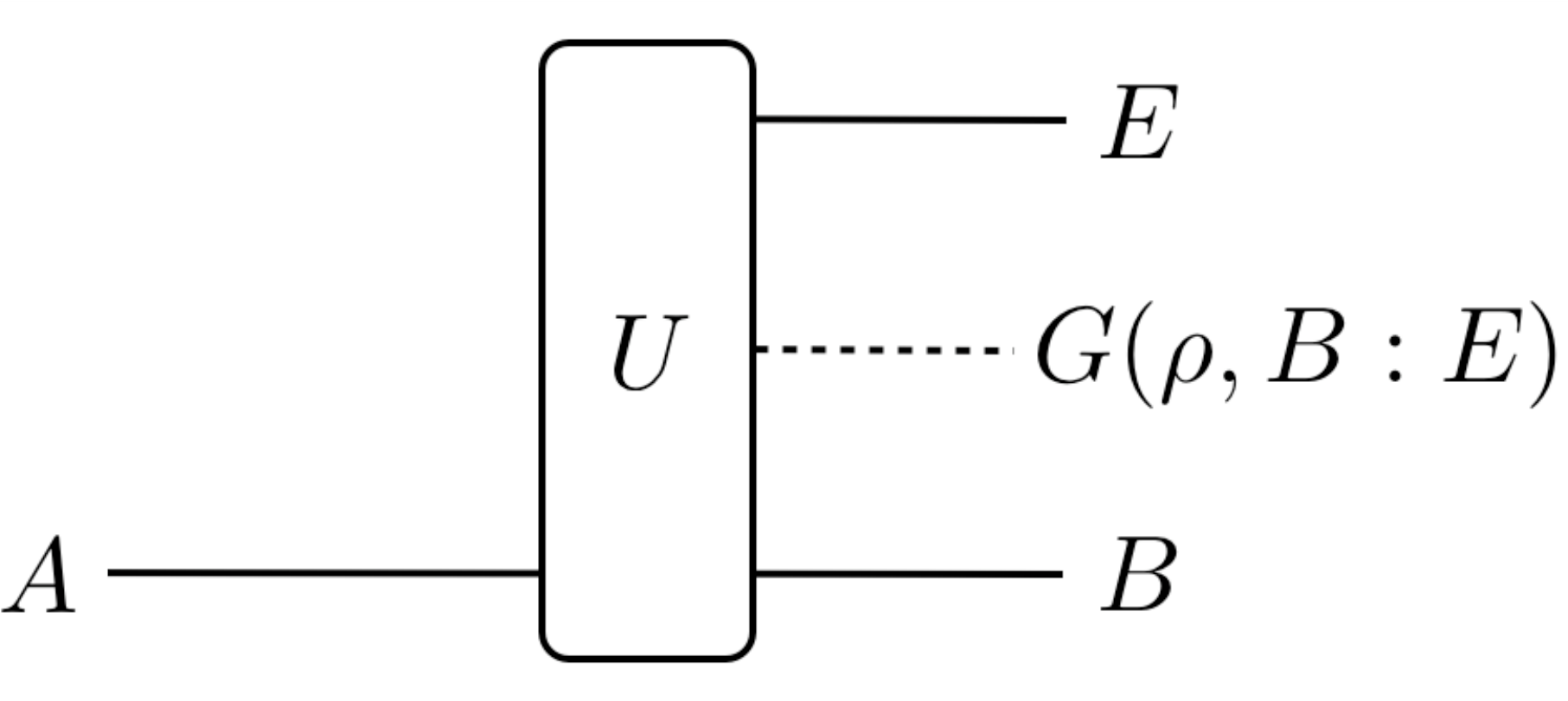}
 \vspace*{-4mm}
  \end{center}
  \caption{Upper bound to the classical information capacity in terms of entanglement measures
  between the output and its environment in a dilation of the quantum channel.}
  \end{figure}

{\it Notation and setting.} We start our discussion by fixing the notation and clarifying some basic concepts that will be used later on.
We consider general quantum channels of
arbitrary finite dimension, $T:{\cal S}(\cc^d)\rightarrow {\cal S}(\cc^d)$,
modeling any general noisy quantum evolution. $T$ is hence an arbitrary trace-preserving completely
positive map. Such a channel can always be written in terms of a {\it Stinespring dilation}
as
\begin{equation}
    T(\rho) = \text{tr}_E(
    U \rho U^\dagger
    ),
\end{equation}
labeling the input by $A$, associated
with the Hilbert space $\cc^d$, the output by $B$ and the environment by $E$,
equipped with Hilbert spaces
$\cc^d$ and $\cc^{d_{\rm env}}$, respectively. $U$ is an isometry
mapping the input on $A$ onto a quantum state on $B$ and $E$.

The classical information capacity, or short
 {\it classical capacity}, of a quantum channel is the rate
 at which one can reliably send classical information.
 It is related to the
{\it Holevo-$\chi$} \cite{Holevo} or the {\it single-shot classical capacity}
of that channel,
\begin{equation}
    \chi(T) = \max\biggl(
    S\bigl(\sum_j p_j T(\rho_j) \bigr)- \sum_j p_j S(T(\rho_j))
     \biggr),
\end{equation}
where the maximum is taken over probability distributions and states,
as the asymptotic regularization
\begin{equation}\label{cc}
    C(T) = \limsup_{n\rightarrow\infty} \frac{\chi(T^{\otimes  n})}{n}.
\end{equation}
The trivial maximum value that the capacity can possibly take is given by the
{\it maximum output entropy} of the quantum channel,
\begin{equation}
    C(T) \leq \max_\rho S(T(\rho))= S_{\max}(T) .
\end{equation}
We will say that whenever this bound is saturated, so when
$C(T)=  S_{\max}(T)$,  that the channel
has {\it maximum capacity}, giving rise to the maximum that is
trivially possible. Of course, this notion
includes the situation
of a {\it perfect quantum channel} that has maximum output entropy of $
C(T)=  S_{\max}(T)= \log_2 (d)$.

The {\it quantum capacity} of a quantum channel, in turn, is related to the
rate at which one can reliably send quantum information through a quantum channel.
Writing
\begin{equation}\label{qc}
    Q_1(T) = \max_{\Psi} \left(S(\omega_B) - S(\omega_E)\right),
\end{equation}
calculated in the state $\omega= |\phi\rangle\langle \phi|$, $\Psi=|\psi\rangle\langle\psi|$,
as a state on $R$, $B$, $E$, and where
$|\phi\rangle := U
    |\psi\rangle$,
with $U $ being again the isometry of $T$, mapping $A$ to $B$ and $E$, and $|\psi\rangle$
being a state vector on $R$ and $A$.  The quantum
capacity is then
\begin{equation}\label{cc}
    Q(T) = \limsup_{n\rightarrow\infty} \frac{Q_1(T^{\otimes  n})}{n},
\end{equation}
again, referred to as {\it maximum} if $Q=S_{\max}(T)$.

{\it Main result.} We can now formulate the main result.

\begin{observation}[Entanglement cannot enhance classical capacity
of  noisy quantum channels to its maximum value]\label{perfect}
Every quantum channel that is noisy (in the sense that the
single-shot classical capacity is not the maximum output entropy) cannot
be made having maximum capacity under the help of any sophisticated
entangled input.
\end{observation}

So if there is a gap to the maximum possible single-shot capacity,
this gap will be preserved in the asymptotic limit, independent of $n$: No entangled input can overcome this
limitation. The single-shot classical capacity may be non-additive, as has been shown in Ref.\ \cite{Hastings}.
Yet, entanglement can only help to some extent, and can, in particular, not make any
imperfect channel perfect.

{\it Upper bounds for classical capacities.}
In order to show this result and the equivalent one for the quantum capacity,
we make use of upper bounds to channel capacities, starting
with the classical capacity. The bounds forming the tools of the argument
will be provided by quantities that capture the entanglement between a system and its
environment in a dilation of the channel. We first show what properties a general
quantity $M: {\cal S}(\cc^d\otimes \cc^{d_{\rm env}})\rightarrow \rr^+$, defined on bipartite
quantum systems, should have.
In order to be entirely clear, we will always give the tensor factors
with respect to which an entanglement measure will be taken. For example, $M(\sigma, A:B)$ would be the quantity evaluated for $\sigma$ with respect to the split $A:B$.
Two properties will be important:
\begin{enumerate}
\item $M$ has the property that
\begin{eqnarray}
    E_F(\sigma, A:B)\geq \sum_{j=1}^n
    M(\sigma_{A_j,B_j}, A_j : B_j)
\end{eqnarray}
for every bipartite state $\sigma$ defined on $n$ copies of a
$d\times e$-dimensional quantum system, labelled $B_1,\dots, B_n$
and $A_1,\dots, A_n$, $\sigma_{A_j,B_j}$ denoting the respective
reduction. \item $M$ is {\it faithful}. That is, $M(\rho, A:B)> 0$
for bipartite states $\rho$ on $A$ and $B$ if and only if $\rho$
is entangled with respect to this split.
\end{enumerate}
Here, $E_F$ denotes the {\it entanglement of formation} \cite{RMP}.
As it turns out, for any quantity satisfying Property 1,
the following bound holds true:

\begin{observation}[Upper bound for the classical capacity]\label{cc}
For any quantum channel $T$ and any
quantity $M$ that satisfies the condition 1.\ we find the single-shot upper bound
\begin{equation}
    C(T) \leq \max_\rho \left( S(T(\rho)) - M(U\rho U^\dagger,B:E) \right).
\end{equation}
\end{observation}

The argument leading to this bound is remarkably simple: Starting
from Eq.\ (\ref{cc}), and defining $\sigma= U^{\otimes  n} \rho
(U^{\otimes  n})^\dagger$ with reductions $\sigma_{B_1,\dots,
B_n}= \text{tr}_{E_1,\dots, E_n} (\sigma)$, $B$ being formed by
$B_1,\dots, B_n$ and $E$ by $E_1,\dots, E_n$, we find, using the
MSW-correspondence \cite{MSW},
\begin{eqnarray}
    \chi(T^{\otimes  n}) &=& \max_{\rho} \left(
    S(T^{\otimes  n}(\rho)) -
     E_F(U^{\otimes  n}
    \rho (U^{\otimes  n})^\dagger,B:E)
    \right)
     \nonumber\\
     &=&
     \max_{\rho} \biggl(
    S(\sigma_{B_1,\dots, B_n}) -
     E_F(\sigma, B:E)
    \biggr)\nonumber\\
     &\leq &\max_{\rho} \biggl(
     \sum_{j=1}^n S(\sigma_{B_j})  -
     E_F(\sigma, B:E)
    \biggr),
\end{eqnarray}
using subadditivity, and hence, using Property 1,
\begin{eqnarray}
    \chi(T^{\otimes  n})
     &\leq &\max_{\rho} \sum_{j=1}^n \biggl(
      S(\sigma_{B_j})  -
     M(\sigma_{B_j,E_j}, B_j: E_j)
    \biggr)
    \nonumber\\
     &\leq &
     n \max_\rho \left( S(T(\rho)) - M(U\rho U^\dagger, B:E) \right),
\end{eqnarray}
which is the above single-shot bound of Observation \ref{cc}.

This bound is to be compared
with the MSW expression \cite{MSW}
for the Holevo-$\chi$ itself,
\begin{equation}
    \chi(T) =\max_{\rho} ( S(T(\rho)) - E_F(U \rho U^\dagger, B:E) ).
\end{equation}
This is very similar, except that now the entanglement of
formation takes the role of the quantity $M$. This indeed leads
also to the conclusion of Observation \ref{perfect} for the
classical capacity: $C(T)$ achieves the maximum upper bound
$\text{max}_{\rho} S(T(\rho))$ if and only if $\chi(T)$ achieves
it. This is because $\chi$ achieves it if and only if
\begin{equation}
    E_F(U\rho U^\dagger, B:E ) = 0
\end{equation}
for the maximizing $\rho$ in $\text{max}_{\rho} S(T(\rho))$,
which means that $U\rho U^\dagger$ has to be separable. Now, if $M$ is also faithful,  i.e., it satisfies Property 2,
then we can see that also $C$ achieves $\text{max}_{\rho} S(T(\rho))$ iff the optimal
$U\rho U^\dagger$ is separable \cite{CostNote}, which proves Observation \ref{perfect}. Below we shall provide a
list of quantities, most of them satisfying both of the postulates.

{\it Identifying candidates for suitable entanglement measures.}
This result, needless to say, leaves the question of finding
entanglement measures exhibiting the above properties 1.\ and 2.
I.e. we need at least one such measure to prove the claim.
Moreover, any computable measure satisfying 1.\  will give rise to
a useful bound for capacity.

{\it (a) The entanglement measure $G$:} Define as in Ref.\ \cite{Henderson}
\begin{eqnarray}
    C_\leftarrow (\rho,B:E)& =& S(\rho_B)\\
    &-&
    \inf \sum_{i=0}^{k-1} q_i S\biggl(\frac{\text{tr}_E((\id\otimes P_i) \rho(\id\otimes P_i)^\dagger)}{q_i}\biggr)
    , \nonumber
\end{eqnarray}
where the infimum is performed over all Kraus operators
$P_0,\dots, P_{k-1}$ acting in $B$ only, satisfying
\begin{equation}
    \sum_{i=0}^{k-1} P_i^\dagger P_i=\id,
\end{equation}
and $q_i= \text{tr}((\id\otimes P_i) \rho(\id\otimes
P_i)^\dagger)$. This is a computable single-shot quantity. We
denote the convex hull of this function with $G$,
\begin{equation}
    G(\rho, B:E) =\min \sum_j p_j C_\leftarrow (\rho_j,B:E),
\end{equation}
where $\rho= \sum_j p_j \rho_j$, and which is an ``entanglement
measure'' in its own right (it is at least a monotone under
one-way LOCC). We claim that this function has the right
properties.

\begin{observation}[Bounding capacities in terms of classical correlations] The quantity $G$ has the properties 1 and 2.
\end{observation}

In fact, the validity of Property 2 is easily shown: Every separable state will have a convex combination
in terms of products, for each of which $C_\leftarrow $ will vanish. In turn, if a state is entangled,
then there must in any convex combination be at least an entangled and hence correlated term, which
will be detected by $C_\leftarrow $. To show Property 1, we can make use of a result of Ref.\ \cite{DongYang}: For a pure
tripartite state $\rho$ shared by $A$, $B$, and $C$, a duality relation gives rise to
\begin{equation}
    S(\rho_A) = E_F(\rho_{A,B},A:B) + C_\leftarrow (\rho_{A,C},A:C).
\end{equation}
We now use the steps of Ref.\ \cite{DongYang} iteratively.
    For a mixed four-partite state $\rho$ on $A$, $B$, $C$, and $D$, the optimal decomposition
    for $E_F(\rho,AB:CD)$ in terms of pure states being $\{p_j,\rho_j\}$, for each $\rho_j$ we have
    $S(\rho_{j;A,B}) = E_F(\rho_{j;A,B,C},AB:C) + C_\leftarrow (\rho_{j;A,B,D},AB:D)
    \geq E_F(\rho_{j;A,C},A:C) +  C_\leftarrow (\rho_{j;B,D},B:D)$.
    Hence
\begin{eqnarray}
    && E_F(\rho,AB:CD) = \sum_j p_j S(\rho_{j;A,B})\nonumber\\
    &\geq& \sum_j p_j \left(E_F(\rho_{j;A,C},A:C) + C_\leftarrow(\rho_{j;B,D},B:D)\right)\nonumber\\
    &\geq&
    E_F(\rho_{A,C},A:C) + G(\rho_{B,D},B:D),
\end{eqnarray}
arriving at Property 1.
This gives rise to a computable bound. Explicitly, it reads
\begin{equation}
    C(T) \leq \max_{(\{p_j\},\{\rho_j\})}
    \biggl( S(T(\rho)) -
     \sum_j p_j C_\leftarrow (U \rho_j U^\dagger ,B:E)
     \biggr),
\end{equation}
with $\rho= \sum_j p_j \rho_j$, as a single maximization. A lower bound to this
is
\begin{equation}\label{SimpleBound}
    C(T) \leq \max_{\rho}
     S(T(\rho)) - \min_\rho
      C_\leftarrow (U \rho U^\dagger ,B:E),
\end{equation}
which is usually less tight, but much simpler to compute.

{\it (b) Variants of the relative entropy of entanglement:} The measure proposed in Ref.\
\cite{Piani} is {\it superadditive} and not larger than the entanglement of formation,
implying Property 1. It is also shown to be faithful
in Ref.\ \cite{Piani}, which is Property 2.

{\it (c) Squashed entanglement:} The {\it squashed entanglement}  \cite{Squashed}
$E_{\rm sq}$  is also known to be
superadditive and is bounded from above by the entanglement of formation, so qualifies
as a bound for the same reason. It is not easily computable, however, as it is based on
a construction involving a state extension the dimension of which is not bounded.
However a lower bound to squashed entanglement was provided  in Ref.\ \cite{B}:
\begin{equation}
    E_{\rm sq}(U\rho U^\dagger, B:E)\geq \frac{1}{4\ln(2) d d_{\rm env} }
    \left(\min_{\sigma}\|U \rho U^\dagger -\sigma \|_1\right)^2,
\end{equation}
in terms of the trace-norm distance to the set of separable quantum states $\sigma$
with respect to the split $B:E$.

{\it (d) Distillable entanglement:} A not efficiently computable but in instances
practical bound is provided by the LOCC or PPT {\it distillable entanglement}
with respect to $B:E$. (Note that either version of distillable entanglement
does not satisfy property 2.)

\begin{figure}
  \begin{center}
 \includegraphics[width=5.5cm]{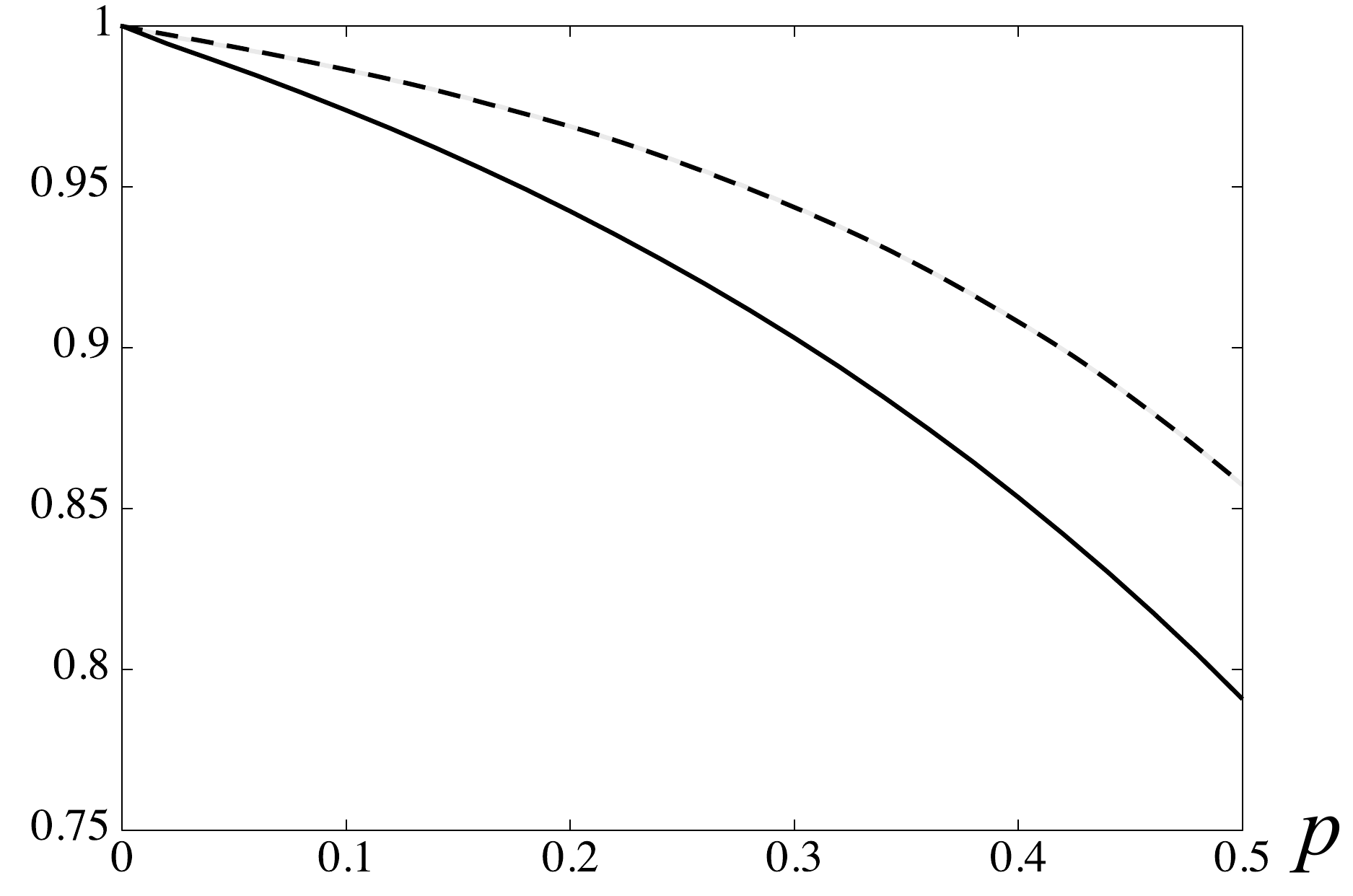}
  \end{center}
  \caption{Upper bound to the classical information capacity
  of the qubit amplitude damping channel as a function of
  $p\in [0,1/2]$. The chosen Kraus operators delivering good bounds for $k=2$ are given by
  $P_0=|0\rangle\langle 0|/2$, $P_1= \sum_{i,j=0,1} |i\rangle\langle j|/x$, $P_2 =
  (\id - P_0^\dagger P_0-P_1^\dagger P_1)^{1/2}$ for $x=4$ (dashed) and $x=3$ (solid). Any choice
  for Kraus operators yields a bound, and with the given choice (determined via numerical
  explorations) one obtains a particularly tight bound.}
  \end{figure}

{\it Example: The amplitude qubit damping channel.} To find any non-trivial bound for the
capacity of the amplitude damping channel has been an open problem for some time \cite{Leung}. The methods
proposed here give rise to such bounds. The Kraus operators of
$T(\rho)= K_0\rho K_0^\dagger + K_1 \rho K_1^\dagger$ are given by
\begin{equation}
    K_0=  \sqrt{p}|0\rangle\langle 1|,\,\,\,\,
    K_1=
    |0\rangle\langle 0| + \sqrt{1-p}|1\rangle\langle 1|
\end{equation}
for $p\in [0,1]$.
The isometry $U$ of this qubit channel maps
\begin{eqnarray}
    |0\rangle\mapsto |0,0\rangle,\,\,\,\,
    |1\rangle\mapsto \sqrt{p} |0,1\rangle + \sqrt{1-p}|1,0\rangle.
\end{eqnarray}
To bound the correlation measure $C_\leftarrow(U\rho U^\dagger,B:E)$, any choice for $k$ and for
$P_0,\dots,P_{k-1}$ giving rise to a positive operator valued measure amounts to a valid bound.
This gives rise to the bound depicted in Fig.\ 2 \cite{Numerics} for $p\in[0,1/2]$.
Note that it is {\it significantly} tighter
than the trivial bound $C(T)\leq  S_{\max}(T)$, which here takes the value $1$.
It is easy to see that for $p\in[0,1/2]$ there always exists an input diagonal in the computational basis
that yields an output ${\rm diag}(1,1)/2$ with unit entropy. For $p=0$ the channel becomes the perfect
channel with $C(T)=1$. (The {\it entanglement assisted classical information capacity} \cite{CE} is also a crude
upper bound,  but yields values even larger than $1$ for $p\in [0,1/2]$).
We have hence established a first
non-trivial bound for the amplitude damping channel.
Needless to say, the same techniques can be
applied to any finite-dimensional quantum channel.

{\it Quantum capacity.} Indeed, an argument very similar to the above one for the classical
capacity of a quantum channel holds true also for the quantum capacity.
We arrive at the following conclusion (for details, see EPAPS \cite{EPAPS}).
Again, entanglement can help to a certain degree, but never uplift channels to the maximum
possible value.

\begin{observation}[Entanglement cannot enhance the quantum capacity to its maximum value]\label{qperfect}
For every quantum channel for which the single-shot quantum capacity is not yet already given
by the trivial upper bound $ S_{\max}(T)$, the same will hold true for the quantum capacity.
\end{observation}

{\it Summary and outlook.} In this work, we have investigated the converse question to the
additivity problem: How much can entanglement help enhance capacities of quantum channels.
In the focus of interest was the question whether entanglement can
ever enhance the capacity to its trivial maximum if a single invocation does not yet
reach that. We affirmatively answer that question to the negative, including the
quantum and classical capacity. In doing so, we have established practical
computable upper bounds to capacities, relating them to
entanglement measures and rendering bounds and witnesses to the latter quantities useful to assess capacities.
There is though an interesting challenge: all the quantities from our list
exhibit a sort of monogamy, i.e., for states which are highly sharable they have to be small,
implying that the bounds may become loose.
An example is a channel whose Stinespring dilation
    gives rise to a $d$-dimensional anti-symmetric space. A normalized projector onto this subspace
    is $d$-sharable, which means that for large $d$ all our bounds would
    tend to $\log d$, while a direct approach of Ref.\ \cite{ChristandlSW-asym}
    shows that the capacity is bounded by a constant independent of $d$.
An open question is therefore how to find a quantity satisfying
our Property 1, but that would not necessarily drop for sharable states.
It is the hope that the present work triggers further work on how
``small'' violations of additivity really are in practice and
what role entanglement plays after all in quantum communication.

{\it Acknowledgements.} We thank M.\ Christandl, A.\ Harrow, M.\
P.\ M{\"u}ller, and A.\ Winter for useful feedback. FB is
supported by a ``Conhecimento Novo'' fellowship from the Brazilian
agency Funda\c{c}\~ao de Amparo a Pesquisa do Estado de Minas
Gerais (FAPEMIG). JE is supported by the EU (QESSENCE, MINOS,
COMPAS), the BMBF (QuOReP), and the EURYI. MH is supported by the
Polish Ministry of Science and Higher Education grant N N202
231937 and by the EU (QESSENCE). DY is supported by NSFC (Grant
10805043). Part of this work was done in the National Quantum
Information Centre of Gda\'nsk. FB, JE and MH thank the
hospitality of the Mittag-Leffler institute, where part of this
work has been done.

\section{EPAPS}

In this supplementary material, we detail the proof of Observation 4 of the main text.
We again label in the Stinespring dilation
$T(\rho) = \text{tr}_E(
    U \rho U^\dagger
    )$ of the quantum
    channel $T$ the input by $A$, the output by $B$ and the environment by $E$, but also keep a system $R$
holding a purification of the input. We have to show that if $Q(T) = S_{\max}(T)$, then also
\begin{eqnarray}
    Q_1(T)= S_{\max}(T).
\end{eqnarray}
If $Q(T) = S_{\max}(T)$ indeed holds true, then
we find for the classical and quantum entanglement-assisted capacities $C_E$ and $Q_E$,
respectively,
\begin{eqnarray}\label{ineq}
    C_E(T) = 2Q_E(T) \geq 2Q(T) = 2S_{\max}(T).
\end{eqnarray}
We also know that
\begin{eqnarray}
    C_E(T) &=& \max_\Psi I(\Psi,  R : B),
\end{eqnarray}
where the maximum is taken over pure states $\Psi$.
One way of expressing the right hand side, so the mutual information,
in our notation of input $A$, output $B$, environment $E$, and
purification $R$ is
\begin{eqnarray}\label{b}
    C_E(T) = \max_\Psi \left(
    S(\rho_R) + S(\rho_B)  - S(\rho)\right),
\end{eqnarray}
where $\rho$ is a mixed state that is shared between $B$ and $R$
that is obtained as $\rho = (\id_R\otimes T)(\Psi)$. From the
subaddivity of the von-Neumann entropy, it follows that for the
input state that achieves the maximum in Eq.\ (\ref{b}),
\begin{equation}
    2S(\rho_B) \geq S(\rho_R)+S(\rho_B) -S(\rho) \geq 2S_{\max}(T).
\end{equation}
Since at the same time, by definition $S(\rho_B)\leq S_{\max}(T)$, we find that there exists an
input pure state $\Psi = |\psi\rangle\langle \psi| $
shared between $R$, $B$, and $E$, with $|\psi\rangle = U|\phi\rangle$,
$|\phi\rangle$ being shared between $R$ and $A$, such that
\begin{eqnarray}
    S(\rho_B) &= & S _{\max}(T),\\
    S(\rho_R)+S(\rho_B) -S(\rho) &=& 2 S_{\max}(T).
\end{eqnarray}
The latter equality also implies that
\begin{eqnarray}
    S(\rho_{B,E})  = S(\rho_B) + S(\rho_E),
\end{eqnarray}
which in turn means
$\rho_{B,E}= \rho_B\otimes \rho_E$.

In the final step, it is the aim to construct an input to the channel that certifies that the single shot
quantum capacity $Q_1(T) = S_{\max}(T)$. Based on the above properties of the channel, this can easily be done.
By virtue of the Schmidt decomposition,
there exists a unitary $V$ supported on $R$ such that
\begin{eqnarray}\label{out}
    V|\psi\rangle &=& |\xi\rangle_{R_1, B} |\eta\rangle_{R_2,E}\nonumber\\
    &=&
    \sum_{i,j}
    \alpha_i \beta_j |i\rangle_{R_1} |e_i\rangle_B |j\rangle_{R_2} |e_{j}\rangle_E,
\end{eqnarray}
where for convenience the tensor factors have been indicated. So the input
\begin{equation}
    V|\phi\rangle = \sum_{i,j}
    \alpha_i \beta_j |i\rangle_{R_1} |j\rangle_{R_2} |d_{i,j}\rangle_A,
\end{equation}
will give rise to the output as in Eq.\ (\ref{out}). From this input to the original problem we can construct a new input which
achieves the desired bound: Take
\begin{eqnarray}
    |\nu\rangle = \sum_i\alpha_i  |i\rangle_{R_1}  |0\rangle_{R_2} |d_{i,0}\rangle_A,
\end{eqnarray}
yielding the output
\begin{eqnarray}
     U|\nu\rangle =  |\xi\rangle_{R_1, B} |0\rangle_{R_2}|e_0\rangle_E,
\end{eqnarray}
which is a tensor product between $R$ and $E$. This means that $Q_1(T) = \max_\Psi ( S(\rho_R) - S(\rho) )= S_{\max}(T)$,
which is that was to be shown.


\begin{thebibliography}{99}
\bibitem{CostNote}
    This result may indeed be viewed as the channel analogue of the
    observation
    that the entanglement cost $E_C$ is strictly positive for every
    entangled state \cite{DongYang}.
\bibitem{Relent}
    With $\rho^{i} = (\id\otimes P_i) \rho(\id\otimes P_i)^\dagger)/{q_i}$, one finds
    \begin{eqnarray}
        C_\leftarrow (\rho,B:E) = \sum_{i=0}^{k-1} q_i S(\rho^{i}_B||\rho_B).
    \end{eqnarray}
    Still, the quantity $C_\leftarrow$ is not convex.
\bibitem{Numerics}
    This is numerically evaluated by optimizing over
    mixed-state ensembles $\rho= \sum_j p_j \rho_j$
    using global optimization both based on the  routine
    ${\tt fmincon}$ in Matlab with randomly sampled initial
    conditions and simulated annealing.
\end{thebibliography}
\end{document}